\documentclass[useAMS,usenatbib]{mn2e}

\usepackage{graphicx}
\usepackage{amssymb}
\usepackage{times}
\begin{document}

\title [Spiral structure in nearby galaxies II] {Spiral structure in nearby galaxies \\ II. comparative analysis and conclusions}
\author[S. Kendall, C. Clarke, R. C. Kennicutt, Jr.]{S. Kendall$^1$
, C. Clarke$^1$\thanks{E-mail:cclarke@ast.cam.ac.uk} ,R. C. Kennicutt$^1$ \\
$^1$ Institute of Astronomy, University of Cambridge, Madingley Road, Cambridge CB3 0HA}
\date{submitted, accepted}

\pagerange{\pageref{firstpage}--\pageref{lastpage}}

\maketitle

\label{firstpage}

\begin{abstract}

 This paper presents a detailed analysis of two-armed spiral structure in a sample of 
galaxies from the Spitzer Infrared Nearby Galaxies Survey (SINGS), with particular 
focus on the relationships between the properties of the spiral pattern in the stellar disc 
and the global structure and environment of the parent galaxies.
Following Paper I we have used a combination of Spitzer Space Telescope mid-infrared 
imaging and visible multi-colour imaging to isolate the spiral pattern in the underlying
stellar discs, and we examine the systematic behaviours of the observed amplitudes and 
shapes (pitch angles) of these spirals.  
In general, spiral morphology is found to correlate only weakly at best with morphological 
parameters such as stellar mass, gas fraction, disc/bulge ratio, and v$_{flat}$. 
In contrast to weak correlations with galaxy structure 
a strong link is found between the strength of the spiral arms and tidal forcing from
nearby companion galaxies.  This appears to support the longstanding suggestion that 
either a tidal interaction or strong bar is a necessary condition for driving grand-design
spiral structure.  The pitch angles of the stellar arms are only loosely correlated 
with the  pitch angles of the corresponding arms traced in gas and young
stars. We find that the strength of the shock in the gas and the contrast in the star formation
rate are strongly correlated with the stellar spiral amplitude.

\end{abstract}


\section{Introduction.}\label{intro}

This paper is the second of two detailing an observational study of spiral structure for galaxies in the \textit{Spitzer} Infrared Nearby Galaxies Survey (SINGS - 
Kennicutt et al. (2003). This work also builds directly on an earlier study of M81 presented in 
Kendall et al. (2008). In the first paper (Kendall et al. 2011, hereafter Paper I), the methods and results from individual galaxies were presented together with a comparison between those  galaxies in which grand design spiral structure
was detected and those that lacked such structure.  In this paper we focus on the subset of 
$13$ galaxies (henceforth the `detailed sample') in which it has been possible to
characterise the properties of the spiral structure in detail and present an analysis of
any trends within this sample. 

Observational results have been crucial to the advances in the study of spiral structure, and the launch of the \textit{Spitzer} Space Telescope provided a new resource in its high quality infrared data. This paper aims to build on previous studies in the optical and infrared, and provides a new dataset against which numerical simulations of spiral structure can be tested.

There have been numerous observational studies of spiral structure, some key results are as follows: 
Schweizer (1976) analysed six galaxies using optical data to produce profiles of surface brightness against azimuth. From these data 
Schweizer concluded that there must be an old disc component in the spiral arms of each of the galaxies studied and that the spiral arms reach amplitudes as large as 30 per cent relative to the disc. Other studies have confirmed this result; the measured ratio of arm to disc mass varies with galaxy and the exact method and wavelength used, but as 
a rule the  contrast between arm and inter-arm regions in the optical 
is in the range 20 to 100 per cent  
(e.g 
Schweizer (1976); Elmegreen \& Elmegreen (1984); Grosbol et
al. (2004)), with more recent studies suggesting that an upper limit of 40 to 50 per cent is appropriate in the majority of cases (note that  contamination from star forming regions can cause this quantity to be over-estimated  particularly in the early studies which used the \textit{I} band rather than near infrared (NIR) data). The amplitude of  spiral arms is potentially  important since it might be
expected to directly affect the global star formation rate (SFR) in
the galaxy: larger contrasts between the stellar arm-interarm regions are expected to cause the gas to shock more strongly, become more dense, and thus increase the star formation (bearing in mind that
according to the Kennicutt-Schmidt law 
(Kennicutt 1998), star formation varies as $\Sigma^{1.4}$.)
This association between star formation and spiral structure would appear to be borne out in cases like M81 and M51, where the vast majority of star formation lies on well defined spiral arms. Some studies have attempted to correlate  the rates of star formation with the strength of the stellar spiral arms: for example, 
Cepa \& Beckman (1990) found that star formation is triggered preferentially by spiral arms (with a `non-linear dependency'). 
Seigar \& James (1998) found a link between SFR measured from far-infrared luminosity and \textit{K} band arm strength, and found that the SFR increases with arm strength. Similarly, 
Seigar \& James (2002) reported that the spiral arm strength (in \textit{K}) correlates with  H$\alpha$ (a SFR tracer) suggesting that stronger potential
variations (and associated shocks)  lead to a larger SFR (up to a limiting threshold, above which the SFR is constant). 
Nevertheless, the picture does not always appear to be this simple:  studies of flocculent galaxies  showing underlying m=2 spiral structure in the mass
density as traced by near infrared emission 
(Elmegreen  et al. 1999, Thornley 1996) clearly show that star formation is not always preferentially located on two well defined spiral arms.  More recently Elmegreen 
et al (2011) analysed NIR imaging for a sample of nearby galaxies from the 
Spitzer S$^4$G survey (Sheth et al. 2010) and found that the distinctions between
grand design and flocculent arm structures seen in the visible extend to the
NIR structure as well.

 Another property of spiral galaxies that has been extensively studied is the  pitch angle (i.e. the angle between spiral features and 
the tangential direction): 
Kennicutt \& Hodge (1982) measured pitch angles and compared them against theoretical predictions from both QSSS (quasi-stationary spiral structure - 
Lindblad 1964; Lin \& Shu 1964,1966) and SSPSF 
(stochastic self-propagating star formation - 
Mueller \& Arnett 1976; Gerola \& Seiden 1978) 
theories and found qualitative   trends that were  in agreement with both
theories but poor detailed agreement in both cases.
Similarly, 
Kennicutt (1981)  compared pitch angle against RSA type (largely dependent on disc resolution), and found a good  correlation on average, but with a large degree of scatter. A correlation between pitch angle, \textit{i}, and bulge-to-disc ratio was also found, where a larger pitch angle (more open spiral
pattern) is associated with a smaller bulge fraction;   again there is a large scatter in the pitch angle at given bulge to disc ratio.  In addition the maximum rotational velocity was found to correlate strongly with pitch angle, with
galaxies with large rotational velocities having more tightly wound spirals.

  However, there are indications that the appearance of spiral arms in the NIR and optical do not always correlate (an extreme case being the optically
flocculent galaxies with NIR arms already noted above).  Some studies using NIR data (e.g. 
Seigar \& James (1998) have not found the same trends as 
Kennicutt. Block \& Wainscoat (1991) found evidence that the spiral arms are less tightly wound in the NIR than the optical. Likewise, 
Block et al. (1994) compare the NIR and optical morphology for a sample of galaxies and, for a number of spirals, found significant differences between the pitch angles of the arms in the optical and NIR. They found that galaxies generally appear to be of an earlier Hubble type when viewed in the NIR (\textit{K'} band) than in the optical, the traditional wavelength for such classifications.
 On the other hand, more recent studies have found a generally good agreement
between the pitch angles measured in the NIR and B band (Seigar et al. 2006, Davis et al. 2012); in the few cases where the two values are significantly 
different, there is a mild preference for the NIR pitch angles to be
larger. Turning now to correlations with galactic parameters, 
Seigar \& James (1998) found no evidence of a correlation between NIR
pitch angle and bulge fraction or Hubble type. Seigar et al. (2005) however
reported a convincing connection between pitch angle in the NIR and
the morphology of the galactic rotation curve, with open arms correlating
with rising rotation curves, while pitch angles decline for flat
and falling rotation curves.

 The goal of this paper is to use the new measurements of the properties of the underlying
{\it stellar} spiral arms from Paper I to re-examine the degrees to which the strengths and 
shapes of these arms are driven by the structural and environmental properties of their
host galaxies.  This is done mainly by examining correlations between the amplitude and 
pitch angles of the stellar arms with measures of galaxy morphology (independent of spiral
structure) and mass, and with a measure of tidal perturbation from nearby companion galaxies.
The same data are also used here to compare the shapes of the stellar arms with those traced
by young gaseous and stellar components, and to constrain the degree to which the response 
of the gas discs and the local star formation rates are influenced by the amplitude of the
spiral perturbation in the stellar discs.

 The data and methods used in this work are described in detail in Paper I. A brief summary is as follows; 3.6 and 4.5$\mu$m NIR data from the \textit{Spitzer} Infrared Array Camera (IRAC) are used in conjunction with complementary optical data in order to trace the stellar mass. In addition, IRAC 8$\mu$m data are used to trace the gas response, with particular emphasis on shocks. The data were analysed using azimuthal profiles (defined from fitting isophotal ellipses to the axisymmetric components of the galaxies) or radial profiles in the case that difficulties were encountered using azimuthal profiles. Azimuthal profile data were further examined by calculating the Fourier components in order to examine the amplitude and phase of individual modes, particularly the m=2 component. This Fourier analysis was not possible for radial profile data; instead the total amplitude and phase were analysed. In this paper we consider only the $13$ galaxies that
constituted the so-called `detailed sample' discussed in Paper I.
We list  various properties of these galaxies (which we go on to
correlate with the properties of their spiral structure) in Table 1.  
 The methods developed to isolate, identify, and parametrise the massive stellar arms are not well suited
to strongly barred galaxies (which can distort arm shapes very significantly from a logarithmic
shape), so this paper focuses mainly on normal and weakly barred galaxies; the role of
bars in spiral structure is another important question awaiting further study
(see Elmegreen et al. 2011).


 \begin{table*}
  \centering
 \begin{tabular}{|l|l|l|l|l|l|l|l|c}
    \hline
    Galaxy & HT & EC & $v_{flat}$ & C & log ($M_*$) & $A/\omega $ & gf   & SSFR \\
    \hline
    NGC 0628  & Sc  & 9  & 217 & 2.95  & 10.1 & 0.39 &  0.32 & $ 6.4 \times 10^{-11} $\\
   NGC 1566  & Sbc   & 12 & 196  & 3.57 & 10.7  & 0.55 & 0.18 & -  \\
    NGC 2403 &  Scd  & 4 & 134 & 3.16  & 9.7  & 0.36  & 0.39 & $7.6 \times 10^{-11}$   \\
    NGC 2841& Sb  & 3  & 302     &  3.27  & 10.8  & 0.48 & 0.17 & $ 1.2 \times 10^{-11}$ \\
   NGC 3031  & Sab & 12  & 229  & 3.91  & 10.3  & 0.58 & 0.04 & $2 \times 10^{-11}$ \\
    NGC 3184& Scd & 9  & 210  & 2.40  & 10.3  & 0.21 & 0.22 & $4.5 \times 10^{-11}$  \\
    NGC 3198 & Sc  & 9  &  150 & 2.54  & 10.1  & 0.21 & 0.51 &  $7.4 \times 10^{-11}$ \\
    NGC 3938 & Sc  & 9 & 196 &  2.95  & 10.1 & 0.41 & 0.39 & $9.5 \times 10^{-11}$  \\
    NGC 4321  & Sbc  & 12 & 222  & 3.03  & 10.9  & 0.36 & 0.21 & 
$6.9 \times 10^{-11}$ \\
    NGC 4579  & Sb & 9  & 288  &  3.92  & 10.9  & 0.41 & 0.06 & $1.1 \times 10^{-11}$ \\
    NGC 5194  & Sc & 12  & 219  & 3.0  & 10.6  & 0.55 & 0.125 & $7.8 \times 10^{-11}$ \\
    NGC 6946 & Sd   &  9   & 186  &  2.83  & 10.5  & 0.27 & 0.25 & $ 1.0 \times 10^{-10}$   \\
    NGC 7793& Sb  & 2  & 115 &  2.48  & 9.5  & 0.22 & 0.29 & $7.4 \times 10^{-11}$      \\
    \hline
 \end {tabular}

\caption{The properties of the galaxies analysed. The columns denote Hubble Type (HT), Elmegreen Class (EC), galactic rotation velocity (in km s$^{-1}$) in flat portion of rotation curve, 
concentration (C),  
logarithm of total stellar mass, 
shear parameter ($A/\omega$) of Seigar et al. 2005 (see Section 2.1.3), gas fraction (gf) and specific star formation rate (SSFR) in yr$^{-1}$. 
$v_{flat}$, log$(M_*)$ , gf and SSFR are mainly derived from Leroy et al. (2008)
as detailed in Sections 2.1.2, 2.1.3 and 2.1.4 respectively. 
C is derived from Bendo et al. (2007) (see Section 2.1.3 for definition).}

\end{table*}

\section{The relationship between galactic properties and the
nature of spiral structure.}\label{ch5_influence}

Previous studies in the optical and infrared have found conflicting trends relating the properties of spiral  features with a range of galactic  parameters such as Hubble type, galaxy mass and rotation curve morphology.  This section sets out the findings from this work and discusses  them in the context of previous results. 

\subsection{Amplitude  and radial extent of spiral structure}\label{ch5_strength}

Figure 1  presents data on the relative  amplitude of the m=2
component at $3.6 \mu$m as a function of radius. 
The axisymmetric components used to normalise the m=2 components contain contributions from the disc and bulge; the halo contribution is not included. In M81, the only galaxy where the halo contribution has been considered  
(Kendall et al. 2008), adding the halo to the axisymmetric components makes very little difference to the relative amplitude, although this is not
necessarily true in all cases. 
Data is presented for the $8$ galaxies in
the `detailed' sample for which it was possible to analyse azimuthal profile
data and thus deduce the amplitude of various Fourier modes.
\begin{figure}   \centering
  \includegraphics[width=83mm]{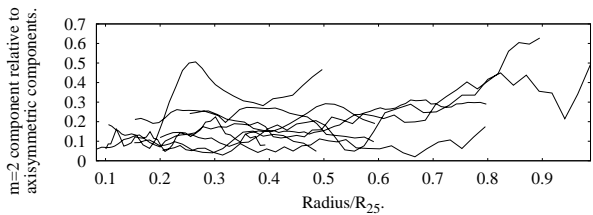}
  \caption{Relative amplitude of the m=2 component of the 3.6$\mu$m (stellar mass) data for all the `grand design' spirals which could be analysed with Fourier components. Note this does not include data obtained from radial profiles. The data are shown for the radial range over which a logarithmic spiral could be traced.}
  \label{m_2_plot}
\end{figure}

 The galaxies with much less power in m=2 are normally those in which there is no single dominant Fourier component and the power in
modes m=1, 3 and 4 
is similar to that in m=2 (these tend to be the optically flocculent galaxies like NGC 7793). 
The galaxy with a smaller radial range than many but a higher relative amplitude is NGC 1566.
\begin{figure}.
  \label{m_2_plot}
\end{figure}

 In what follows we represent the `average' amplitude of spiral structure
in each waveband and for each galaxy as a straight average of  the values
obtained over the range of radii for which a logarithmic spiral pattern
is detectable.  These averages are listed for each galaxy in Table 2. Most galaxies thus have amplitudes in three bands
(optical, IRAC1 and IRAC2) which are denoted by cross, dot and star
symbols in each case. The data for each galaxy are linked by vertical lines.
The larger symbols and dashed vertical lines correspond to the galaxies
whose structure has been determined through radial profile analysis.
The largest  crosses  denote the  optical data for the three galaxies
(NGC 3031,NGC 4321 and NGC 5194) which are apparently undergoing the
strongest tidal interactions (i.e. with $P > 10^{-3}$ (equation 1); see
Table 4 and Figure 17).
 In almost all cases the main uncertainty is associated with
the range of values at different radii (see Figure 1) and the 
issue of how datapoints at small and large radii should be weighted.
The difference in mean amplitude between different wavebands is mainly
due to the fact that in some galaxies the radial range over which
a logarithmic spiral can be traced is different in different bands.
The difference between amplitudes in different wavebands is therefore
a measure of the uncertainty associated with radial averaging within
the radial range studied (which is generally less than $R_{25}$; see
Figure 1).
 Note that in the case of galaxies that overlap with the
sample of Elmegreen et al. (2011), the amplitude of variation at given
radius is in good agreement between the two papers (once account is
taken of the difference in the definition of arm contrast); the average
values quoted in the Elmegreen et al. study are however generally
higher due to the fact that the spiral structure is generally analysed out
to larger radius.  As noted above, in the bulk of our  plots the vertical lines
link average values in different wavebands; in order to avoid a confusing
plethora of vertical lines 
we show the errors associated with radial averaging (as fine vertical lines)
only in a single plot
(Figure 4).  
The measurement errors for {\it individual}  datapoints at given radius are generally much smaller
than the uncertainties associated with radial averaging, being
much less than 0.1 magnitudes in most of the azimuthal profile galaxies: see
individual profiles and associated errors in Paper I.

 \begin{table}
  \centering
 \begin{tabular}{|l|l|l|l|l|l|l}
    \hline
    Galaxy &$\Delta_V$  &$\Delta_{3.6}$& $\Delta_{4.5}$ &  $\Delta \phi_v$ & 
$\Delta \phi_{3.6}$& $\Delta \phi_{4.5}$\\
    \hline
    NGC 0628  & 0.131 & 0.198  & 0.187 & 6.25 & 8.33 & 8.22 \\
    NGC 1566  & 0.247 & 0.284 & 0.287 & 3.92 & 4.50 & 4.57 \\
    NGC 2403 & - & 0.119 & 0.130 & 3.89 & 3.89 & 3.83   \\
    NGC 2841& -  & 0.069 & 0.076    &  2.85 & 2.85 & 3.60 \\
   NGC 3031  & 0.260& 0.224 & 0.207  & 1.79 & 1.89 & 2.19 \\
    NGC 3184& 0.225 & 0.282 & 0.304 & 6.57 & 5.55  & 5.49  \\
    NGC 3198 & 0.131 & 0.213 &  0.200 & 2.98  & 5.69 & 5.42 \\
    NGC 3938 & 0.091 & 0.093 & 0.086 & 5.01 & 4.49 & 4.37 \\
    NGC 4321    & 0.284 & 0.327  & 0.351 & 4.59 & 4.46 & 4.19 \\
    NGC 4579  & 0.177 & 0.188 & 0.166 &  3.06 & 3.50 & 3.76 \\
    NGC 5194  & 0.299 & 0.416  & 0.425 & 4.89 & 5.96 & 5.28 \\
    NGC 6946 & 0.166 & 0.214  & 0.222 & 1.97 & 1.56 & 1.55  \\
    NGC 7793& 0.085 & 0.071  & 0.077 &  4.78 & 4.93 & 4.97 \\
    \hline
 \end {tabular}
\caption{The parameters of spiral structure in the visible, $3.6 \mu$m (IRAC1)
and $4.5 \mu$m (IRAC2) bands: $\Delta$ is the mean spiral amplitude (see text)
and $\Delta \phi$ is the total angle subtended by the arms in radians}

\end{table}

\subsubsection{Dependence on Hubble type and Elmegreen type.}
 We first plot the amplitudes of spiral structure as a function of
Hubble type and Elmegreen type in Figures 2 and 3. There is no obvious
correlation with Hubble type unless one excludes the three galaxies of
type Sb and earlier. 
The correlation between spiral amplitude and Elmegreen class is
much more evident: this is unsurprising since Elmegreen class is based
purely on spiral morphology (albeit in the optical) with 
patchy, wispy spirals corresponding to arm class 1-4 and classes
5-12 corresponding to galaxies with increasingly prominent two armed
grand design spiral structure. The fact that this classification scheme
tracks the NIR amplitudes is a confirmation that
the over-all level of spiral structure from galaxy to galaxy is
reasonably well correlated in the optical and the NIR.
We also note that the three `strongly interacting galaxies' are associated
with high Elmegreen class (i.e. well defined grand design structure).

These results are broadly consistent with those from an independent study
by Elmegreen et al. (2011), based on Spitzer S$^4$G observations.  In particular
both studies show the absence of any correlation between arm amplitude and
Hubble type when all spiral arm types are considered together, though
Elmegreen et al.  (2011) do see evidence for a possible correlation between
arm-interarm contrast and Hubble type when flocculent galaxies are considered
separately.  Our observation of a correlation between arm amplitude and 
Elmegreen spiral type is also consistent with a trend observed in the
Elmegreen et al.  (2011) sample.

\begin{figure}   
\centering
  \includegraphics[width=83mm]{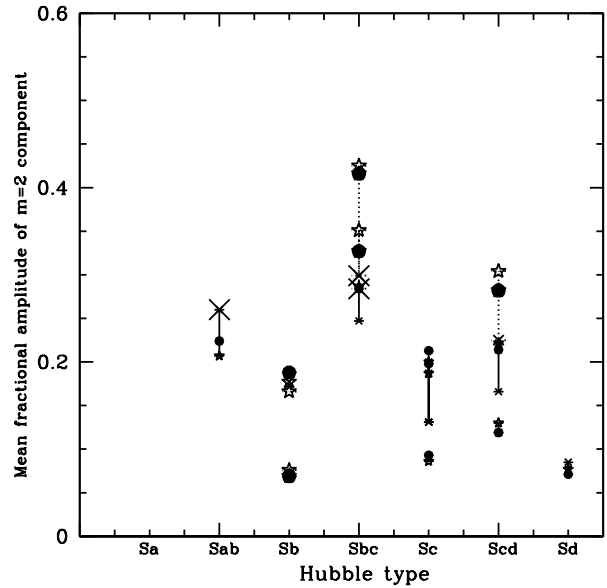}
  \caption{Average m=2 relative amplitude plotted against Hubble type. 
The smaller symbols correspond to the $8$  galaxies analysed with the azimuthal profile method, while the larger symbols denote the $5$  galaxies analysed with radial profiles. Each galaxy has three data points shown, with separate amplitude 
estimates from 3.6$\mu$m, 4.5$\mu$m and optical colour corrected data being
denoted by dots, stars and crosses respectively; datapoints for a given galaxy are linked
with vertical lines  (dashed in the case of the radial profile galaxies).
 Note
that for two galaxies ( NGC 2403 and NGC 2841)  no optical data are   
presented since these two galaxies were too flocculent in the optical
to be able to extract reliable spiral parameters (see Paper I). The largest crosses denote optical data for the three galaxies with $P > 10^{-3}$ (equation
(1)) that are apparently undergoing tidal interactions (see Figure 17 and Table 4). }

  \label{hub_av2}
\end{figure}

 \begin{figure}
\centering
  \includegraphics[width=83mm]{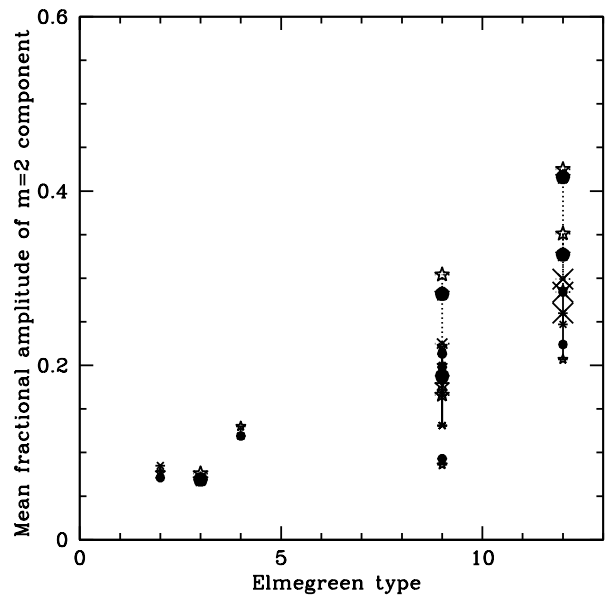}
\caption{Average m=2 relative amplitude plotted against Elmegreen class (see
Figure 2 for explanation of symbols).}
\end{figure}

\begin{figure}
\centering
\includegraphics[width=83mm]{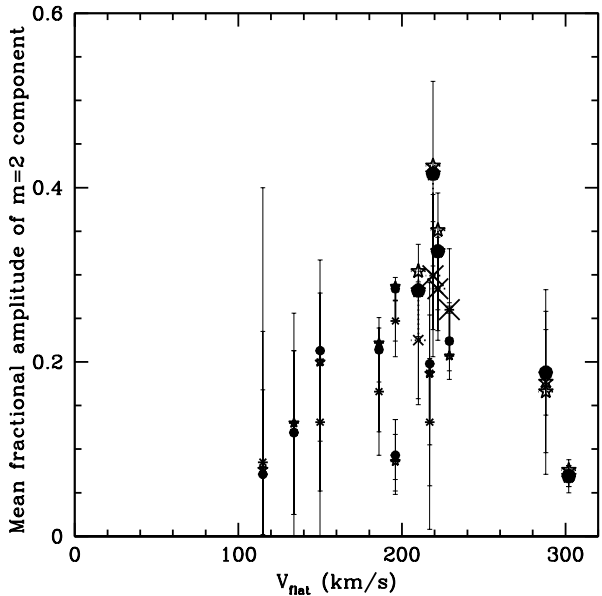}
\caption{Average m=2 relative amplitude plotted against magnitude of rotation 
velocity in the flat region of the rotation curve (see
Figure 2 for explanation of symbols).The fine vertical lines (shown only in this plot for clarity) indicate the errorbars associated with radial averaging
of the spiral amplitudes. }
\end{figure}

\begin{figure}
\centering
\includegraphics[width=83mm]{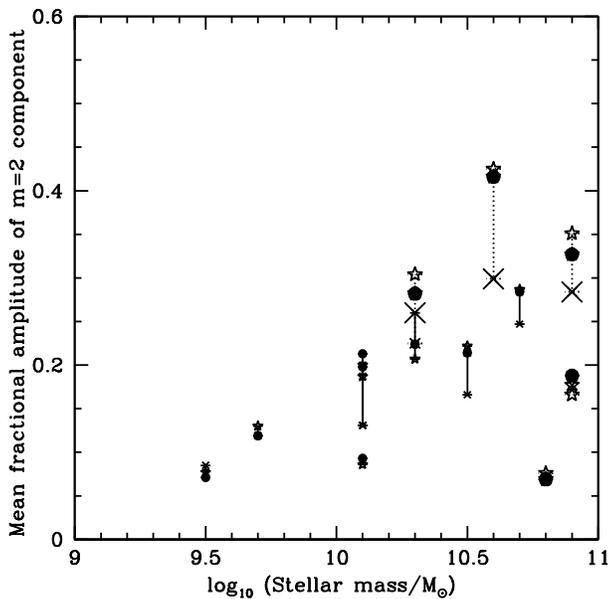}
\caption{Average m=2 relative amplitude plotted against galactic stellar mass
(see
Figure 2 for explanation of symbols).}
\end{figure}

\begin{figure}
\centering
  \includegraphics[width=83mm]{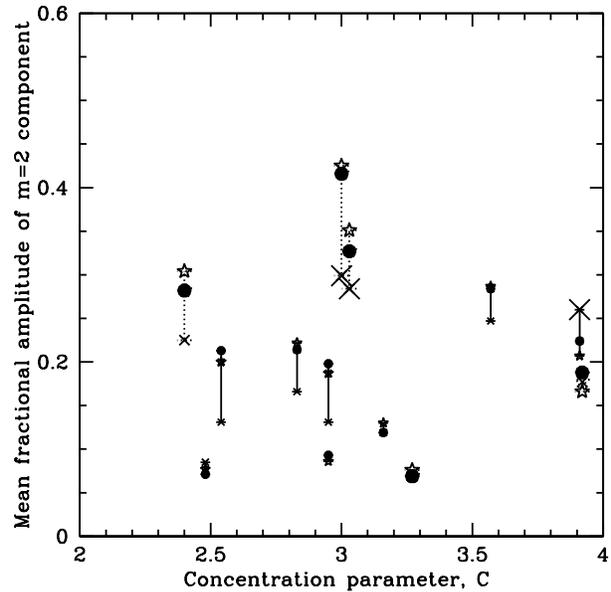}
\caption{Average m=2 relative amplitude plotted against concentration index (see
Figure 2 for explanation of symbols).}
\end{figure}

\begin{figure}   \centering
\includegraphics[width=83mm]{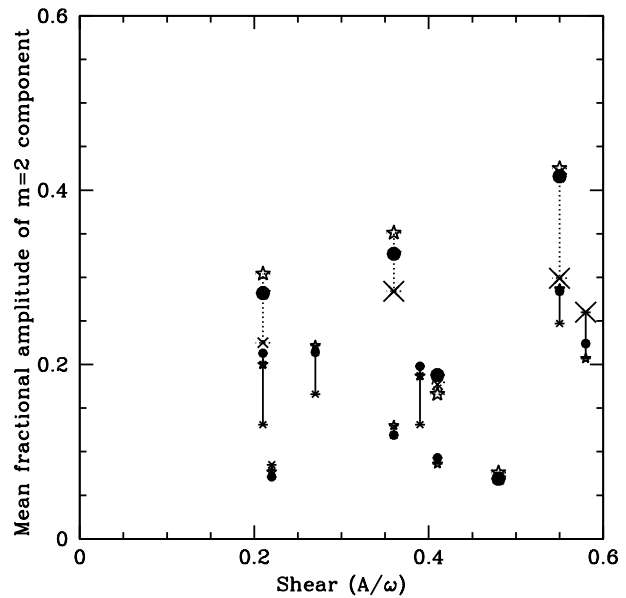}
\caption{Dimensionaless shear (see text) plotted against average m=2 amplitude.
Solid body rotation, flat rotation curves and falling
rotation curves correspond to $A/\omega$ values of $0,0.5$ and $>0.5$
respectively.
   See Figure 2 for explanation of symbols.}
  \label{conC_av2}
\end{figure}

\begin{figure}   \centering
  \includegraphics[width=83mm]{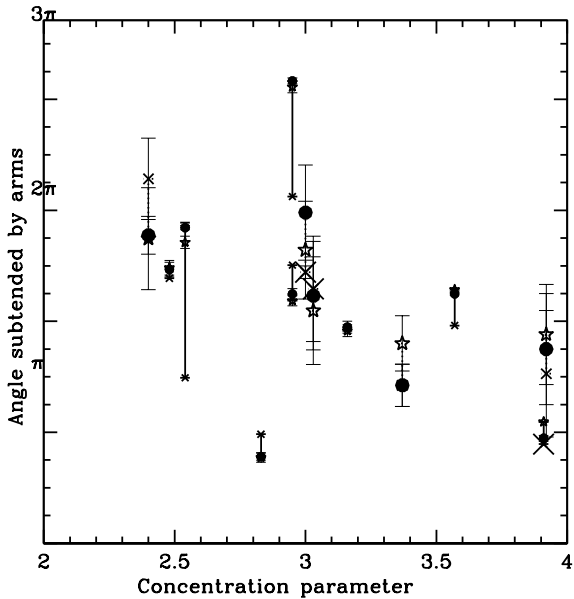}
  \caption{The concentration parameter plotted against angle subtended by the spiral arms (defined within the radial range of the logarithmic spiral).
See Figure 2 for explanation of symbols.}
\end{figure}

\begin{figure}   \centering
  \includegraphics[width=83mm]{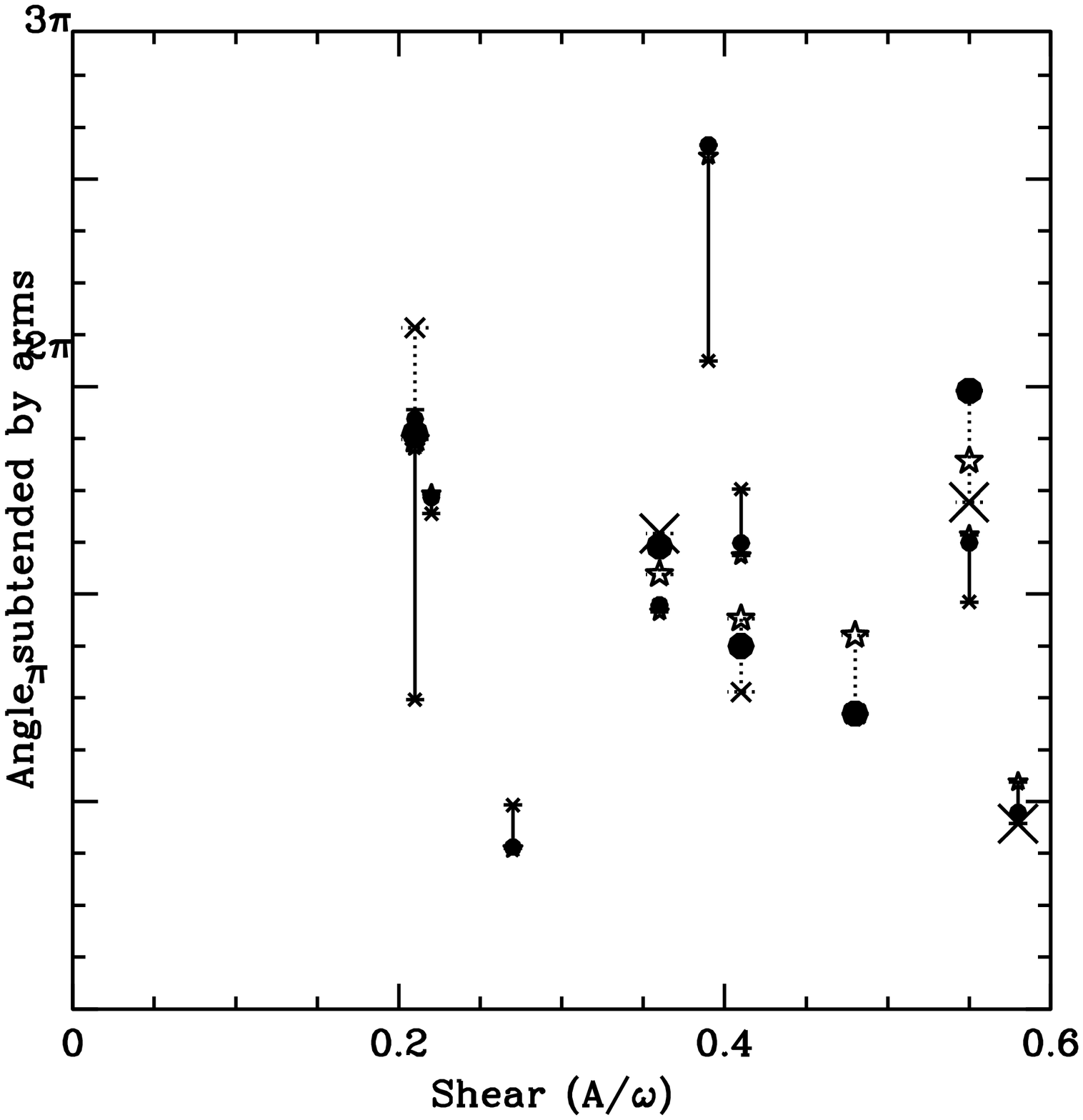}
  \caption{Dimensionless shear (see text)  plotted against angle subtended by the spiral arms. See Figure 2 for explanation of symbols.
}
\end{figure}

 \subsubsection{ Dependence on galactic stellar mass and rotational
velocity}

  Figures 4 and 5 plot the m=2 amplitude against the galactic rotation velocity 
evaluated in the flat region of the rotation curve, $v_{flat}$,  and the stellar
mass $M_*$, with both these quantities being derived, where possible, from
Leroy et al. (2008). Leroy et al. calculate $v_{flat}$ by fitting
the rotation curve with a function of the form $v_{rot}(r) =
v_{flat} (1 - exp (-r/r_{flat})$ where $v_{flat}$ and $r_{flat}$ are free 
parameters. For galaxies not included in the Leroy et al.  sample, the same
function is fit to the kinematic data from the sources given in Paper I,
with errors estimated at the $10$ per cent level. The stellar masses are
calculated by integrating the galactic models produced by GALFIT (Peng et al. 
2002) and applying the prescription to convert to stellar
mass given in Leroy et al. (2008). 

In both plots there is a clear trend for m=2 average amplitude to increase
with stellar rotation velocity and stellar mass, particularly if one
excludes the three galaxies of Hubble type Sb or
earlier. It should not be surprising that the trend is roughly the same
in both plots since the Tully -Fisher relation (Tully \& Fisher 1977)
implies that $v_{flat}$ is an approximate tracer of stellar mass. The same
approximate trend was found by Elmegreen \& Elmegreen (1987) who demonstrated
that spiral arm amplitude is an increasing function of galaxy `size'
(proxied by the product of $R_{25}$ and $v_{flat}$). This trend is
consistent with the decline in spiral amplitudes towards later
Hubble types seen in Figure 2 since the 
general relationship between Hubble type and luminosity implies that
later type galaxies are generally less massive.   A qualitative correlation
between the strength of spiral structure and parent galaxy luminosity has been 
recognised for decades (van den Bergh 1960a, b), and forms the basis of the
van den Bergh luminosity classification of galaxies.  The trends seen in 
Figures 4 and 5 can be regarded as a quantitative manifestation of this 
correlation.  Figure 4 is particularly suggestive in that $11/13$ of the galaxies appear to broadly follow a trend of increasing
amplitude of spiral features with increasing rotation velocity; the low
amplitude of spiral features for the two objects with the largest rotational
velocities (the Sb galaxies NGC 4579 and NGC 2841) are notably
discrepant. We however caution against the drawing of any but the most 
tentative conclusions from this small sample.

\subsubsection{Dependence on concentration and galactic shear}
 Another factor which may influence the amplitude of spiral structure is
the degree of central concentration of galaxies inasmuch as this affects
the morphology of the galactic rotation curve. Kormendy \& Norman (1979)
argued  that spiral waves are damped at the Inner Lindblad Resonance
(ILR) and thus predicted that spiral structure (other than that
driven by companions or bars) should be  restricted to
regions of galaxies with  nearly solid body rotation
(for which there is no  ILR). Their optical data showed some support for this
hypothesis although  this may be partly driven 
by the fact that 
low shear conditions
favour more open spirals (which are more readily identified
observationally). On the other hand this association  between
spiral structure and solid body rotation was {\it not} confirmed by the
infrared study of Seigar et al. (2003) who found that spiral structure
extends well beyond the region of solid body rotation. 
On a theoretical level, Sellwood \& Carlberg (2014) have argued against the significance of
ILRs for wave damping and indeed report a diminished amplitude
of spiral structure in regions of low shear 
(Sellwood \& Carlberg 1984), ascribing this result
to the relative ineffectiveness
of swing amplification in conditions of low shear (Toomre 1981).

   Here we investigate
possible dependences on the rotation curve in two ways. We plot
the amplitude of spiral structure against galaxy concentration parameter
(C) in Figure 6. We use the 
IRAC 3.6$\mu$m concentration index (C) data from 
Bendo et al. (2007),
which is defined as
the ratio of the radius containing  80 per cent of the light to 
that containing 20  per cent of the light, i.e.  $\frac{r_{80}}{r_{20}}$. 
Higher  values of C imply  more concentrated  emission and hence presumably
also mass; the dynamical influence of a strong central mass concentration
is to promote differential rotation (angular velocity declining with
radius) and thus - according to the Kormendy \& Norman argument - might
be expected to be associated with lower amplitude spiral structure.
Figure 6  however shows no evidence of an obvious  (anti-)correlation
between spiral amplitude and concentration index. 

 Figure 7 
investigates the dependence of the strength of spiral structure on
rotation curve morphology more directly, by evaluating for
each galaxy the {\it dimensionless shear} defined by 
Seigar et al. (2005) as:
\begin{math}
\frac{A}{\omega} = \frac{1}{2}(1 - \frac{R}{V}\frac{dV}{dR})
\end{math}
where A is Oort's constant,   $\omega$ is the angular velocity and $V$
the local rotational velocity. In each case this quantity is evaluated for
the region of the galaxy over which a spiral structure is detected. Since 
$A/\omega$ is equal to $-0.5 \times$ the power law index for the
dependence of angular
velocity on radius, it follows that solid
body rotation corresponds to dimensionless shear of zero, a flat rotation
curve to $0.5$ and a falling rotation curve to larger values
of $\frac{A}{\omega}$. The
Kormendy \& Norman hypothesis would preferentially associate
prominent spiral structure with low values of the shear but Figure
7 shows no evidence for this.

The amplitude of the m=2 component is only one measure of the strength of the spiral structure; another is the extent (radial or azimuthal) of the spiral arms. 
Seigar \& James (1998) found  `a hint of a deficit' of galaxies with 
a large bulge to disc ratio  and extended spiral arms, 
quantifying the latter in terms of   the angle subtended by the arms.
  Figure 8  shows the angle subtended by the spiral arms plotted against concentration parameter for this sample (with the fine vertical
liens representing the errorbars on the angle subtended), and provides some  support
for the 
Seigar \& James result. However, if the angle subtended by the arms is plotted against dimensionless  shear as shown in Figure 9, the trend is much less convincing, suggesting that disc shear is in fact not an important contributory factor to spiral arm extent. The mild (anti-)
correlation between spiral arm extent and concentration parameter
would then need an alternative explanation.

  In summary then, we find no relationship between spiral arm amplitude
or  angle subtended by the arms with dimensionless shear and no
relationship between spiral arm amplitude and concentration parameter.
There is arguably a weak anti-correlation between the angle subtended
by the arms and concentration parameter but this does not obviously
relate to the rotation curve morphology.

\begin{figure}   \centering
  \includegraphics[width=83mm]{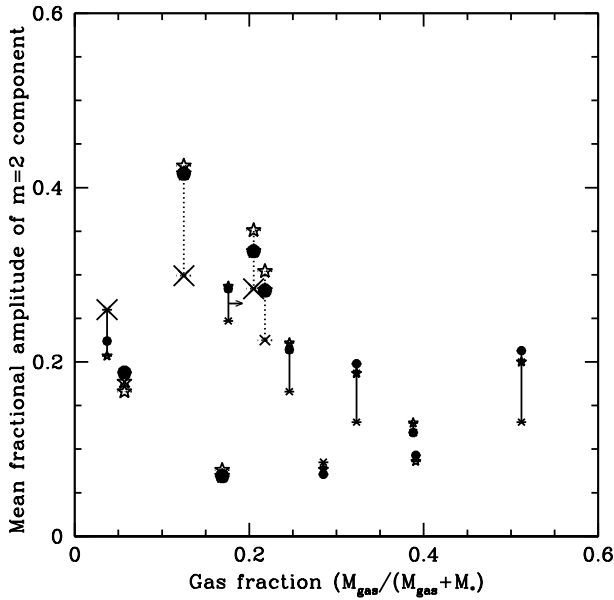}
  \caption{Gas fraction, (M$_{HI + H_{2}}$)/(M$_{HI + H_{2}}$+ M$_{*}$),  against average m=2 relative amplitude.  The arrow indicates that the gas fraction data for NGC 1566 give a lower limit (see  Figure 2 for explanation of symbols ).}
  \label{gasfrac_av2}
\end{figure}



 \subsubsection{Dependence on gas fraction and specific star formation rate}

The gas fraction is plotted against the relative amplitude of the m=2 Fourier component in Figure 10.
Where galaxies are not found in 
Leroy et al. (2008) the HI and H$_2$ masses are taken from 
Kennicutt et al. (2003). For galaxies that are common to both datasets there is a scatter of up to 0.5 dex (but generally more like 0.2 dex) in gas masses. H$_2$ masses are not given for NGC 1566 and NGC 3031: in these cases the total gas mass plotted is in fact only the HI mass. NGC 3031 has very little detected CO, and so this should not be significant, but for NGC 1566 the difference may be larger:
we indicate the fact that the gas fraction is a lower limit in this object
via the arrow in Figure 10.

  Figure 10 suggests a trend of decreasing amplitude of
spiral arms with gas fraction (with again the three earliest type galaxies
having discrepantly low spiral arm amplitude for their gas fraction values).
This trend is in apparent contradiction to the notion that spiral
structure is maintained by dynamical cooling provided by the
continued production of stars from a dynamically cold gas reservoir.
We also note that the three strongly interacting galaxies
each lie at the top of the range of spiral amplitude at given gas fraction.
 
 In Figure 11  we plot the m=2 amplitude against specific star formation (normalised by the stellar mass). SFR data come from 
Leroy et al.  (2008) except in the cases of NGC 3031 and NGC 1566: in NGC 3031 the
SFR data derive from 
Perez-Gonzalez et al. (2006) and references therein. No SFR data are available for NGC 1566.
We note that in principle a positive correlation between spiral arm
amplitude and specific star formation rate could be driven by 
incomplete removal of young stellar emission from the NIR images.

  However  Figure 11 does not suggest any correlations between these
quantities (in contrast to the mild positive correlation seen by
Seigar \& James 1998). This apparent lack of evidence for spiral
structure enhancing the global SFR is discussed further in Section 4.2.
 


\begin{figure}   \centering
  \includegraphics[width=83mm]{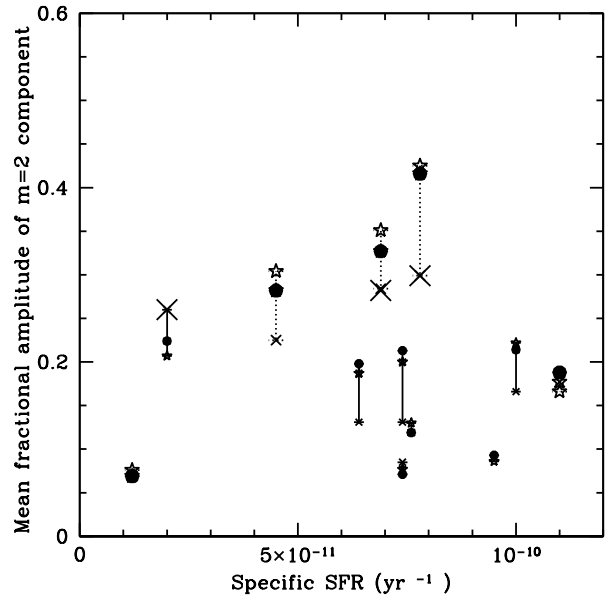}
  \caption{Average m=2 amplitude plotted against specific star formation rate (yr$^{-1}$). See Figure 2 for explanation of symbols. }
  \label{spsfr_av2}
\end{figure}

\subsection{Pitch angles}\label{ch5_pitch}

The pitch angles are calculated by fitting a straight line to the data in the $\phi$ vs ln(R) plots presented in Paper I. The data from each galaxy and wavelength are presented in Table 3.

\begin{table}
  \centering
  \begin{tabular}{|l|l|l|l|l|}
    \hline
    Galaxy & \multicolumn{3}{c|}{Pitch angle (3.6$\mu$m, 4.5$\mu$m, optical)} & method \\
    \hline
    NGC 0628&16.2$_{16.2}^{16.3}$&16.4$_{16.3}^{16.5}$&17.3$_{17.3}^{17.3}$&A\\
    NGC 0628&15.5$_{14.6}^{16.5}$&15.7$_{14.8}^{16.7}$&15.3$_{14.2}^{16.5}$&R\\
    NGC 1566&19.7$_{19.6}^{19.7}$&19.4$_{19.3}^{19.4}$&22.3$_{22.2}^{22.4}$&A\\
    NGC 2403&19.6$_{19.3}^{19.8}$&19.9$_{19.6}^{20.1}$& - &A\\
    NGC 2841&6.7$_{ 6.3}^{ 7.2}$&6.5$_{ 6.1}^{ 7.0}$&-&R\\
    NGC 2841*&9.4$_{ 8.5}^{10.4}$&8.1$_{ 7.5}^{ 8.8}$&-&R\\
    NGC 3031&23.6$_{23.3}^{23.9}$&20.7$_{20.7}^{20.7}$&22.6$_{22.6}^{22.7}$&A\\
    NGC 3184&19.5$_{18.9}^{20.0}$&19.9$_{18.5}^{21.6}$&18.2$_{17.3}^{19.2}$&R\\
    NGC 3198&15.8$_{15.7}^{15.8}$&15.5$_{15.4}^{15.7}$&18.2$_{18.2}^{18.2}$&A\\
    NGC 3938&15.0$_{14.9}^{15.2}$&15.4$_{15.2}^{15.5}$&17.8$_{17.8}^{17.8}$&A\\
    NGC 4321&21.3$_{19.4}^{23.7}$&22.2$_{20.1}^{24.8}$&16.3$_{14.8}^{18.0}$&R\\
    NGC 4579&20.3$_{17.9}^{23.3}$&16.6$_{14.9}^{18.7}$&17.9$_{15.2}^{21.6}$&R\\
    NGC 5194&13.7$_{12.8}^{14.7}$&13.6$_{12.6}^{14.8}$&13.6$_{13.5}^{13.8}$&R\\
    NGC 6946&29.3$_{28.8}^{29.8}$&29.5$_{28.8}^{30.2}$&24.0$_{23.5}^{24.6}$&A\\
    NGC 7793&15.7$_{15.5}^{15.9}$&15.6$_{15.4}^{15.8}$&16.2$_{16.1}^{16.3}$&A\\
    \hline
  \end {tabular}

  \caption{Measured pitch angles at $3.6 \mu$m, $4.5 \mu$m and optical (V band). Method; this column lists the methods used to analyse the galaxy; A= azimuthal profiles, R=radial profiles. It is worth noting that the upper and lower
bounds quoted in the Table result from accounting for  statistical errors; the systematic errors (discussed in the text) may be larger, especially for highly inclined galaxies. *NGC 2841; the first entry gives the pitch angle calculated for R$>$0.45R$_{25}$, the second entry for R$<$0.45R$_{25}$ (the data used in the figures are those for R$\ge$0.45R$_{25}$).}

\end{table}

 Possibly the first point that is worth commenting on is that, as was seen in Paper I, all these galaxies have phase-radius relationships that are close to logarithmic in nature ($\phi \propto ln(R)$) over at least part of the disc of the galaxy. Although logarithmic spiral arms are predicted by QSSS theories (and subsequent related theories such as global spiral modes) this does not automatically lead to the conclusion that spiral arms must obey these relationships; that they do is notable. 
It is unlikely that this is merely a selection bias since the radial
ranges for detectable spiral structure quoted throughout this work are not {\it defined} as regions of
logarithmic spiral structure, even though this turns out to be the
case in practice.
     There is one galaxy, NGC 2841 (pitch angle $\sim$7$^o$), for which the pitch angle should be treated with an extra degree of caution, because as described in Paper I the high inclination and complex structure make the extraction of
a single pitch angle problematical (see caption to Table 3).
  The errors quoted for all galaxies in Table  3 are the errors associated with accurately determining the phase of the m=2 component in the Fourier fits or phase determined from radial profiles; these errors do not take into account the uncertainties in the galaxies' ellipticity or position angle. It is difficult to quantify the associated uncertainties in the pitch angle due to inaccuracies in galaxy orientation, although 
Block et al. (1999) found that by varying the inclination 
and position angle used to fit a highly inclined galaxy (i$\ge$60$^o$) the pitch
angle could change by as much as 10 per cent. Nevertheless, it is striking that
the pitch angles  in the two infrared bands and optical (V) band are in general so similar to each other, with the differences being similar to the
errors quoted in Table 3.

 It is worth considering whether there is a systematic variation in pitch angle measured from the azimuthal profile and radial profiles. If anything, the radial method should produce pitch angles that are systematically smaller, due to measuring more of the contamination from young populations which are expected to be more tightly wound. The only galaxy for which a direct comparison is available is NGC 0628: the small difference ($1 \deg$) is
within the errors. 

\subsubsection{Comparison with previous  (H$\alpha$ and K band ) determinations}
 We now compare the pitch angles found in this work to results from optical (H $\alpha$)  data (Figure 12) for the set of $9$ galaxies
which overlap our sample and that of Kennicutt (1981).  It can be seen that the late type spirals agree well, but that there is a systematic difference for the early types, Sab and Sb. 
The two galaxies that are most discrepant (NGC 3031 (Sab) and NGC 4579 (Sb))
are notably those that were  shown in Paper I to exhibit the 
 most convincing {\it offsets} between gas shocks and stellar 
density maxima. The sense of the offset (gas shock precedes stellar density 
maximum inward of corotation) combined with the fact that one would
expect star formation to follow the shock after a fixed time (and therefore
at an angular offset that decreases with radius for a differentially
rotating galaxy) are both such as to explain smaller pitch angles in
the  H$\alpha$ emission  (assuming this traces recent star formation).
 Our sample size does not allow us to say whether this effect is
a general feature of early type galaxies.

 We can also compare some of our optical pitch angle determinations with
those reported in the literature (see  
Ma 2001 and Davis et al. 2012: $7$ members of our sample have B band  
pitch angle determinations also listed in either or both of these
studies). As reported in the latter paper (Table 3)
the agreement between our results and these other determinations is 
generally good (to within a few degrees or better in most cases).
 
 Turning now to previous pitch angle determinations in the NIR we
note that
there is no overlap in sources between
the galaxies listed in Table 3 and the galaxies previously studied in the
K band by Seigar \& James (1998) and Seigar et al. (2005). These two
previous K band studies report rather different pitch angle distributions:
Seigar \& James 1998 record low values 
(generally in the 
 $5-10 \deg$ range) while
the sample of Seigar et al. is broadly distributed in pitch 
angle, including a number of objects with $i > 30 \deg$.
In that respect the latter distribution is similar to
that found in the B and I band by  Davis et al. (2012). Our
own results follow a much narrower distribution 
( mainly lying in the range $15-20 \deg$ in the optical as well as
NIR bands). This result is most likely a consequence of our small
sample size.

\begin{figure}   \centering 
  \includegraphics[width=60mm]{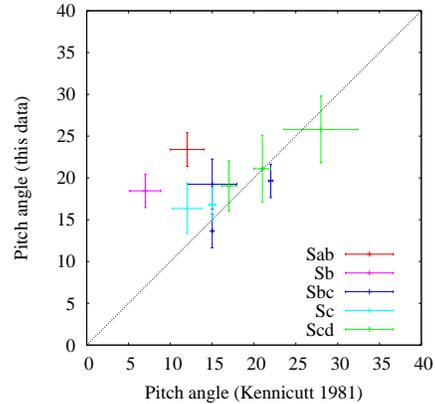} 
  \caption{Pitch angles plotted against data from 
Kennicutt (1981) obtained using H$\alpha$ data. The pitch angles of late type spirals are consistent (to within the errors). Early type spirals have systematically larger pitch angles measured in the NIR.}
  \label{pitch_compare} 
\end{figure}

\begin{figure}   \centering
  \includegraphics[width=83mm]{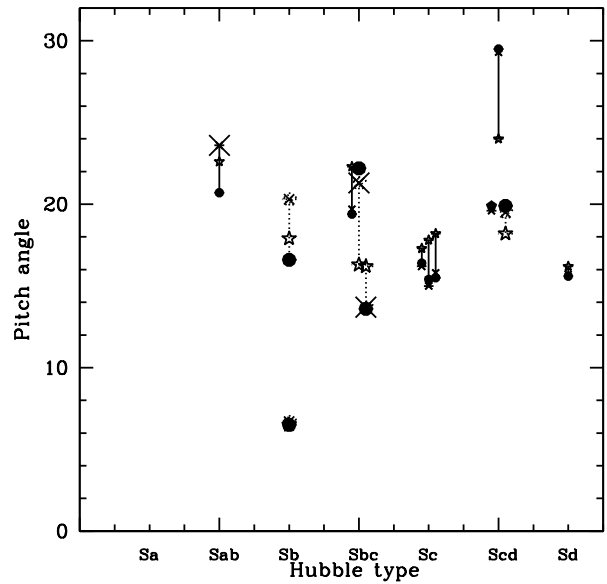}
  \caption{Pitch angles plotted against Hubble type. See Figure 2 for an explana
tion of the symbols}
  \label{hub_pitch}
\end{figure}

\begin{figure}   \centering
  \includegraphics[width=83mm]{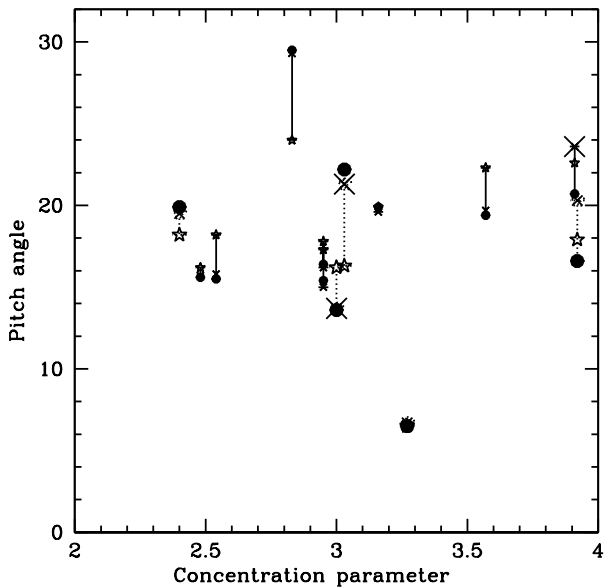}
  \caption{Pitch angles plotted against mass concentration index C.
See Figure 2 for an explanation of the symbols.}
  \label{conC_pitch}
\end{figure}

\begin{figure}   \centering
  \includegraphics[width=83mm]{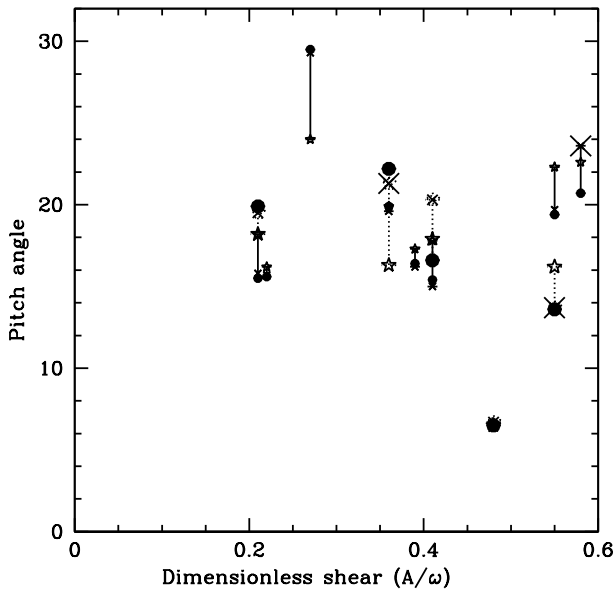}
  \caption{Pitch angles plotted against the dimensionless shear parameter.
See Figure 2 for explanation of symbols.}
  \label{hub_pitch2}
\end{figure}

\begin{figure}   \centering
  \includegraphics[width=83mm]{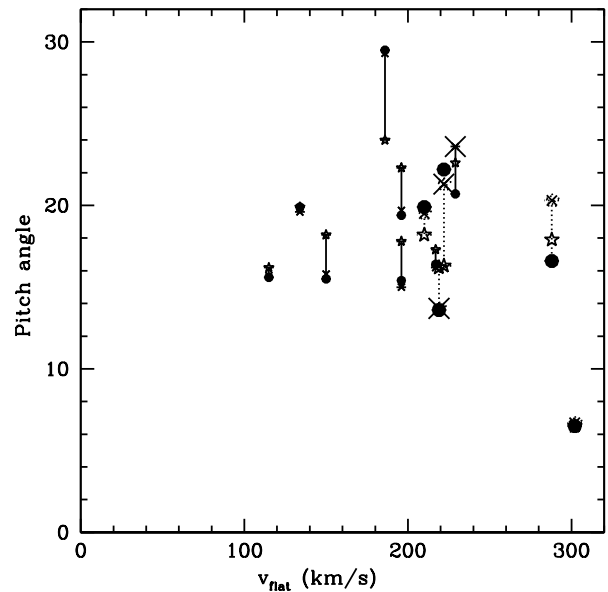}
\caption{Pitch angles plotted against the magnitude of the rotation
velocity in the flat portion of the rotation curve. See Figure 2 for explanation of symbols.}
\end{figure}


\subsubsection{Dependence on Hubble type}
Figure 13  plots the pitch angle of the m=2 components from galaxies in the detailed sample against Hubble type. Roberts et al. (1975) proposed
a relationship between Hubble type and pitch angle under the assumption that
the latter is fundamentally a measure of bulge to disc ratio and using the
dispersion relation of Lin \& Shu (1964) to quantify the expectation that
more loosely wound patterns are to be found in systems that are more
disc dominated. The appearance of Figure 13 is strongly
driven by the two galaxies with unusually large and small pitch angles in 
our sample (NGC 6941 and NGC 2841) and the fact that the former is
of later Hubble type creates an impression of a correlation. However,
the small range of measured pitch angles for the rest of the sample undermines
the case for a correlation. Note that previous 
NIR results
( Seigar \& James 1998) also found no clear evidence of
such a correlation (and again for the reason that the bulk of objects in
the sample were concentrated over a narrow range of pitch angles). Moreover 
the $H \alpha$ study of  Kennicutt (1981) found  only  a weak relationship between Hubble
type and pitch angle ( which is surprising given
that the Hubble classification system implicitly considers the pitch angle of the spiral arms).

\subsubsection{Dependence on rotation curve morphology and concentration.}
  Figure 14  shows likewise that there is no obvious relationship between
pitch angles and  central concentration , again contradicting  the expectation that disc dominated systems should be associated with more open spiral patterns. 
This 
is consistent with the NIR results of Seigar \& James (1998)
and also with Kennicutt (1981) who found only a weak correlation with
bulge to disc ratio. In Figure 15  we plot pitch angles against the 
dimensionless shear parameter of Seigar et al. (2005). 
These data show some weak evidence for more open
structures in the case of galaxies with strongly rising rotation
curves (although this impression is strongly driven by the two 
galaxies with outlying pitch angle values, NGC 2841 and NGC 6946). 
The relationship between pitch angles and galactic shear is much
less evident than in the study 
of  Seigar et al. (2005) and in particular we do not find the
large pitch angles ($i > 30$ deg.) at low shear ($< 0.4$) that
were reported in this work. An association between shear and pitch angle
(in the sense found by Seigar et al. ) is expected theoretically 
(Lin \& Shu 1964, Bertin \& Lin 1996, Fuchs 2000, Baba et al. 2013) and
has recently been explored numerically by Grand et al. (2013) and
Michikoshi \& Kokubo (2014): these simulations show
much larger pitch angles in the case of systems with rising rotation
curves than we find here.

 Finally, we plot in Figure 16 the dependence of pitch angle upon the
galactic rotation velocity in the flat portion of the rotation curve
(see Section 2.1.2). The appearance of an anti-correlation is strongly
driven by one galaxy (NGC 2841) which combines a small pitch angle with
a large rotational velocity, though as noted above the pitch angle
in this case is particularly uncertain due to 
its large inclination and complex structure (Paper I). We thus
find no evidence for the strong anti-correlation between pitch angle and
$v_{flat}$ that was found in the optical by Kennicutt \& Hodge (1982). 

\subsubsection{Summary}
 The results presented in this section show no strong correlations between
pitch angles and other galaxy parameters. The presence or absence
of apparent correlations in Figures 13 -16  need to be interpreted in the
context of the fact that $11/13$ of our galaxies have pitch angles within
a rather narrow range, with only two galaxies having values that are
either $< 10 \deg$ or considerably greater than $20 \deg$ (as such,
our sample is almost all consistent with the predictions of
swing amplification theory which predicts maximum pitch angles
in the range $15-20 \deg$;  Toomre 1981, Oh et al. 2008). However,  as
discussed in Section 2.2.1 above, we believe that this narrow
range reflects the small sample size rather than differences
in analysis method or the different range of wavebands employed
compared with previous studies. Nevertheless the strongly
bunched distribution means that the  placing
of the  two most discrepant galaxies in each plot is a strong driver
of whether there is 
an apparent correlation or not.
With this caveat, we find no evidence to support the association between
large pitch angles and rising rotation curves found by Seigar et al.
(2005) in the near infrared, nor between large pitch angles low rotation
velocities found by Kennicutt \& Hodge (1982) in the optical.
 
We have also
drawn attention to  the suggestion that early type galaxies may be
more tightly wound in H$\alpha$  than in the near infrared.
Finally,  it is evident that the three strongly interacting galaxies do not occupy a distinctive region in Figures 13-16: clearly spiral structure that is driven by galaxy interactions cannot be distinguished from that in isolated galaxies on the grounds of near infrared pitch angle. 
%
\begin{figure}   \centering
  \label{hub_pitch2}
\end{figure}

\section{The relationship between galactic environment and the nature of spiral
structure}

  In the preceding plots we differentiated the three galaxies which we judge to
be undergoing the strongest tidal interactions. Here we present the analysis
of galactic environment that led us to this conclusion and examine
how the strength of tidal interaction affects the amplitude of spiral
structure.

 We used the NASA/IPAC Extragalactic Database to search for companions to the $13$ galaxies considered in this paper.
The companions identified 
 are detailed in  Table 4  
and all lie within
within $\pm$400kms$^{-1}$ of the target galaxy.
 The mass ratio was
calculated using the relative \textit{B} band magnitudes and assumed a constant mass
-to-light ratio.
 Byrd \& Howard (1992) proposed the tidal parameter $P$ as a measure of the
strength of galactic interactions:

\begin{equation}
P = \frac{M_{c}/M_{g}}{(r/R)^{3}}
\label{ch5_eq1}
\end{equation}
 where  $M_{g}$ is the galaxy mass, $M_{c}$ the companion mass, $R$ the galaxy radius, and $r$ is the distance of closest approach. Byrd \& Howard suggested a  minimum value for P of 0.01 and 0.03 for pro- and retro-grade interactions respectively to induce a spiral response to the centre of the galaxy. In
Paper I we analysed the incidence of `strong interactions' (corresponding
to galaxies with $P$ values exceeding a given threshold) in the
sample of galaxies that did and did not exhibit grand design spiral structure.
We  found that this incidence was only different (and higher in grand design
spirals) if a threshold value of $P = 0.01$ was adopted, thus confirming
the estimate of Byrd \& Howard regarding the value of $P$ required for
significant tidal perturbation of the galaxy.

 Whereas in Paper I we assessed companions down to a limiting magnitude
of $B=15$ and such that they lay  within 10 scalelengths
( $\frac{R_{proj}}{R_{25}}$ $<$ 10.0) of the main galaxy, this criterion did
not correspond to a fixed $P$ value (the $6$ galaxies in our sample
that had companions conforming to this criterion are marked with a $+$
symbol  in
Table 4). In the present paper our focus is
to discover whether ({\it within the sample of grand design spirals})
there is a significant correlation between the tidal parameter $P$
and the amplitude of spiral structure. Accordingly we examine all the
companions in the correct velocity range lying within $20$ scale lengths
and evaluate the maximum $P$ value that is contributed by any of the
companions. The mass ratios, normalised distances and $P$ values are
listed in Table 4 and $P$ is plotted against spiral
amplitude in Figure 17. We see that most of the sample have $P$
values of around $10^{-5}$ and indeed we found in Paper I that such
values are also typical of the sample in which grand design spiral
structure was not detected. We also note that a difficulty in relating the observed $P$ value to
the importance of tidal interactions: while  the projected distance may be
a substantial under-estimate of the three-dimensional separation, spiral structure is likely to reflect the strength of the interaction at a previous pericentre passage and thus the projected separation is in this sense an over-estimate of the relevant value.

\begin{table}
  \centering
 \begin{tabular}{|l|l|l|l|l}
    \hline
    Galaxy & Companion & $\frac{R_{proj}}{R_{25}}$ & Relative mass & P\\
    \hline
    NGC 0628 + & UGC 01176 & 9.6 & 0.02 & $2.2 \times 10^{-5}$\\
    NGC 1566 + & NGC 1581 & 9.7 & 0.05 & $5.5  \times 10^{-5}$ \\
    NGC 2403 &NGC 2366 & 20.  & 0.1 & $1.3 \times 10^{-5}$  \\
    NGC 2841& UGC 4932 & 6.1 & 0.01    &  $4.4 \times 10^{-5}$ \\
    NGC 3031 + & M82 & 2.7 & 0.28 & $1.4 \times 10^{-2}$ \\
    NGC 3184& KHG 1013+414 & 7.4 & 0.01  & $2.4 \times 10^{-5}$  \\
    NGC 3198 &* & * & * & $5.5 \times  10^{-5}$ \\
     NGC 3938&-&-&-&-\\
    NGC 4321 +   & NGC 4323 & 1.4 & 0.01 & $3.4 \times 10^{-3}$\\ 
    NGC 4579 + & NGC 4564 & 9.8 & 0.25 & $2.6 \times 10^{-4}$\\ 
    NGC 5194 + & NGC 5195 & 0.8 & 0.25 & $4.9 \times 10^{-1}$\\
    NGC 6946& -& - &- & -\\
    NGC 7793& - & - & - & -  \\
    \hline
 \end {tabular}

\caption{ Values of $P$ together with list of companions, mass ratios and  normalised separations taken from the
literature  for all galaxies with a companion
contributing $ P > 10^{-5}$ (the $6$ galaxies marked with a $+$ sign are
those that were deemed to be interacting according to the `inclusive'
definition of Paper I: see text for details). * In the case of
NGC 3198 there are two galaxies (SDSS J 101848.78+452137.1 and VV 834 NEDOR)
that contribute nearly equal $P$ values and the quoted value
is the sum of these.}
 \end{table}



\begin{figure}   \centering
  \includegraphics[width=83mm]{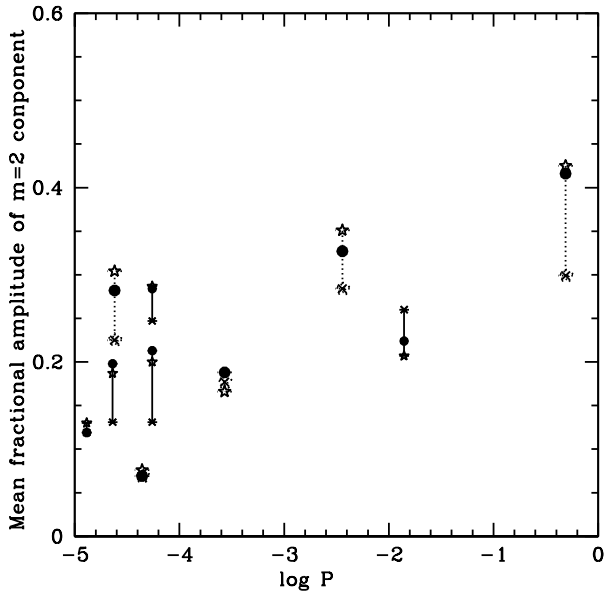}
  \caption{P (see equation (1) plotted against m=2 relative amplitude. P gives a measure of the tidal interaction between two galaxies, and 
correlates strongly with the strength of the spiral. The galaxies with the largest P values ( NGC 5194, NGC 3031 and NGC 4321 in descending order) are
marked with the large crosses  in the other Figures.}
  \label{P_av2}
\end{figure}

  Despite these  caveats, Figure 17 demonstrates that the amplitude of spiral 
responses does indeed correlate positively with the tidal $P$ parameter (in fact
Figure 17 shows a more convincing correlation than any of the other
plots we present here). The fact that $3/13$ galaxies have $P < 10^{-5}$ and
yet display a significant spiral pattern implies that tidal interaction
{\it is clearly not a pre-requisite} for inducing spiral structure 
(and indeed we have seen in Paper I that at least half the sample
of galaxies without detected grand design structure have $P$ values of
$10^{-5}$ or above). On the
other hand, Figure 17 presents convincing evidence that where there is strong
interaction the {\it amplitude of spiral structure is enhanced}. We have
seen that the pitch angles of such `induced' spirals are indistinguishable from
spiral patterns in apparently isolated galaxies. We have also seen that
where we have tentatively identified  trends in the amplitude of spiral features
(i.e. the positive correlations with $v_{flat}$ and stellar mass shown in
Figures 4 and 5 and the mild anti-correlation with gas content shown in Figure 10) then the strongly interacting galaxies (i.e. those with $P > 10^{-3}$) 
roughly follow these trends but at enhanced amplitude. Once again, we
however caution against over-interpretation given our limited sample size. 

 \begin{figure}   \centering
  \label{hubb_m3m2}
\end{figure}

\begin{figure}   \centering
\end{figure}

\begin{figure}   \centering
  \label{hub_pitch2}

\end{figure}

\begin{figure}   \centering
  \label{rotgrad_amp2}
\end{figure}

\begin{figure}   \centering
\end{figure}

\begin{figure}   \centering
\end{figure}

\section{The response to stellar spiral structure: gaseous shocks and associated star formation.}\label{ch5_responses}

The first half of this paper studied the effects of galaxy type, morphology and environment on the stellar spiral structure. However, as well as being influenced by the galaxy, a global  spiral is expected to influence the galaxy in which it is located. The following section considers the effect of the spiral on the 
structure of the gas and the distribution of star formation in galaxies.

\subsection{Gas response}\label{ch5_gas}

\subsubsection{Azimuthal offsets in the location of spiral shocks}

  Here we briefly recapitulate the results of Paper I with regard to
the location of shocks in the gaseous component (as traced by $8 \mu$m
emission) with respect to the spiral in the stellar mass density (as traced
at $3.6 \mu$m).  Of the $13$ galaxies in the detailed sample considered
here, $4$  galaxies (NGC 2403, NGC 2841,
NGC 6946, NGC 7793) exhibited a flocculent response at $8 \mu$m 
so it is not possible to talk meaningfully about the azimuthal alignment
between the spirals manifest in the gas and the stars.  Amongst the
remaining $9$ galaxies there is a general tendency for the $8 \mu$m
spiral to be located on the trailing (concave) side of the stellar ($3.6 \mu$m)
spiral, i.e. (at radii inwards of corotation) being  {\it upstream}
of the stellar spiral. Moreover   the $8 \mu$m spiral is generally
somewhat more
tightly wound. This is most clearly demonstrated in the case of NGC 3031 (M81) 
where the azimuthal offset increased systematically with radius and whose
behaviour was the subject of detailed analysis in Kendall et al. 2008. Four
other galaxies (NGC 3184, NGC 3198, NGC 3938 and NGC 4579) also demonstrated
evidence of a systematic increase of offset angle with radius. In the
remaining cases, the scatter in azimuthal offset is sufficiently large to
prevent the identification of an obvious radial trend. 

  The possible significance of an upstream azimuthal offset between the shock in the gas and the maximum stellar density is discussed in Kendall et al. 2008.  A scenario that predicts such an offset (with magnitude increasing with radius) is the case of a rigidly rotating spiral mode  whose lifetime is sufficiently long to permit the gas to achieve a steady state flow in the imposed potential. 
In this case  the solutions of Roberts (1969) and Shu et al.  (1972)
predict that the gas shocks upstream of the gas inwards of corotation, with the
magnitude of this offset increasing with radius, a result that has been verified through hydrodynamic simulations (Gittins \& Clarke 2004). Kendall et al.  (2008)
used these results to infer the corotation radius in NGC 3031 and obtained a value that is in reasonable accordance with other estimates in the literature.
On the other hand, Clarke \& Gittins (2006) and Dobbs \& Bonnell (2008) 
undertook hydrodynamic simulations of gas in the case that the stellar potential
derived from a high resolution re-simulation of the N-body calculations reported
in Sellwood \& Carlberg (1984).       
Notably in these simulations with a `live' galactic potential, the gas shocks
tended to trace the regions of instantaneous maximum stellar density 
and exhibited  no systematic azimuthal offset with respect to the
stellar spiral. A clear difference between this simulation
and the idealised case involving a single long lived spiral mode is that, 
in the N-body calculations, spiral features come and go on a timescale of a few
orbits. Clarke \& Gittins related  the gas morphology  in this case to the
fact that the gas did not have time to adjust to a steady state configuration
on the timescale  on which individual spiral features established and
dissolved. In the case of spiral structure induced by a strong
galactic encounter (as in the modeling of M51 - NGC 5194 - by Dobbs et al. 
2010) no azimuthal offset was found between the gas and stellar spirals.

  How do these theoretical results relate to the range of results that
we find in our sample? We can divide our sample (as detailed above) into
$4$ galaxies with flocculent $8 \mu$m structure, $5$ with evidence for
a systematic upstream offset with magnitude broadly increasing with 
radius and the remaining $4$ as having no simple offset pattern (we denote
these groups F(locculent), R(egular) and C(omplex): see Paper I for
the phase diagrams on which this classification is based. We find no
systematic differences between the galactic properties in these three
groups apart from the fact that all the galaxies in the F(locculent) category
are isolated ($P < 10^{-5}$; see Table 2); the R(egular) and C(omplex)
galaxies occupy a wide range of $P$ values, although notably the galaxy
which is undergoing by far the strongest interaction (NGC 5194) is in the
C(omplex) category, in line with the simulations of this encounter by
Dobbs et al. (2010).    

  We conclude from this that many of the isolated galaxies  exhibit the
kind of gas response seen in simulations based on live galactic potentials
and 
that the most strongly interacting galaxy also agrees with simulations inasmuch
as it exhibits no clear offset;  the R(egular) group however exhibit
a trend suggestive of a more long lived spiral pattern. We re-emphasise
that this group is heterogeneous in terms of its galactic properties. To date
there has been no detailed quantitative study of how tightly the offset
data can be used to constrain the lifetime and nature of spiral structure
in simulations.

\subsubsection{The strength of the shock as traced by $8 \mu$m emission}
In addition to the position of the shock, it is instructive to 
look at the strength of the gas shocks as a function of the 
strength of the stellar spiral, since a larger spiral amplitude is expected to 
trigger a larger shock. We compared these properties  for all galaxies 
in the sample, not just those for which the offset work was carried out. 
In contrast to  the stellar mass maps, the gas response cannot be normalised by an axisymmetric component, so instead we simply  divide the 8$\mu$m 
azimuthal profile into `on' and `off' arm regions; `on' arm is defined 
as being angles $\pm\frac{\pi}{4}$ of the m=2 peaks (`off' arm is the other $\pi$ degrees, in two arcs). The response is calculated by taking the ratio of the highest peaks in the `on' arm region to the average `off' arm flux.
Note that we use azimuthal profile data for analysis of the
$8 \mu$m emission for all galaxies in the sample, including those whose
amplitude at $3.6 \mu$m is obtained through radial profile analysis.

\begin{figure}   \centering
  \includegraphics[width=83mm]{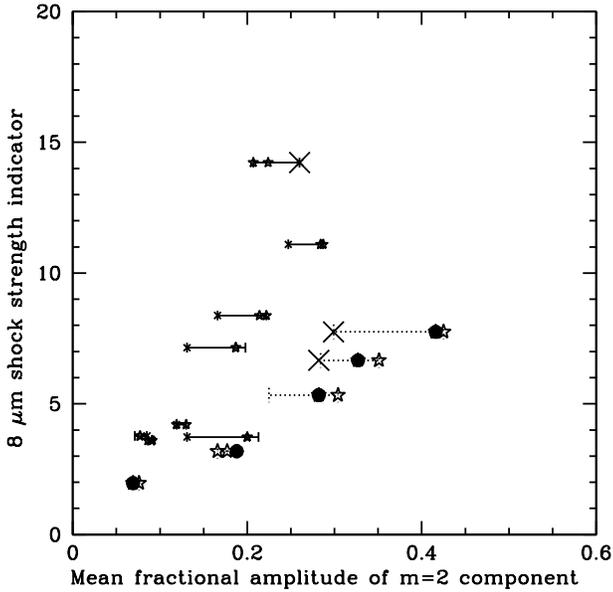}
  \caption{The  $8 \mu$m gas shock strength indicator 
(defined as the average of the 
four highest peaks in the 8$\mu$m profile in the `on' arm region, normalised by 
the average value of the 8$\mu$m profile in the `off' arm region) plotted against average m=2 amplitude
at $3.6 \mu$m.  See Figure 2 for an explanation of the symbols.}
  \label{shockresp_m2}
\end{figure}
\begin{figure}   \centering
  \includegraphics[width=83mm]{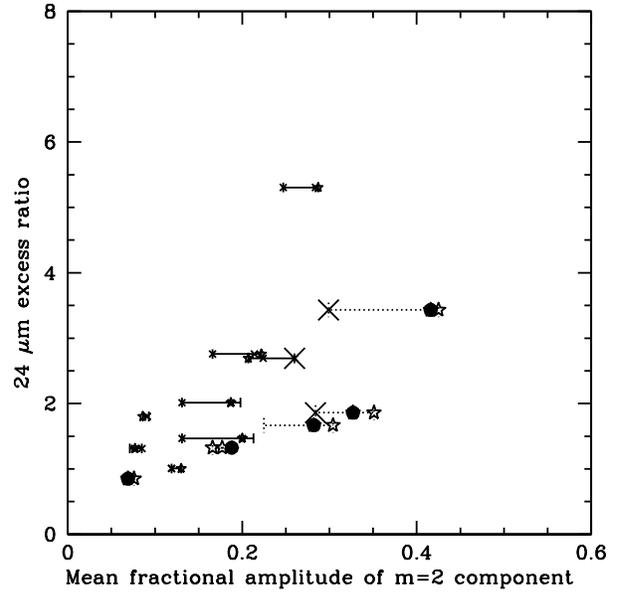}
  \caption{The  $24 \mu$m excess ratio  (see text for definition).
See Figure 2 for an explanation of the symbols.}
  \label{shockresp_m2}
\end{figure}

\begin{figure}   \centering
\end{figure}

Figure 18 
shows the trends in the gas response as a function of m=2 amplitude
and show that, as expected, the gas response will tend to be stronger for a larger stellar wave amplitude. 
(Note that we expect such a correlation to be weakened somewhat by
the fact that $8 \mu$m PAH features may be depleted by the strong UV
emission associated with the enhanced star formation in shocked gas.)

\begin{figure}   \centering
\end{figure}

\subsection{Star formation.}\label{ch5_SFR}

\begin{figure}   \centering
\end{figure}

  In this section we apply  the same method as in Section 4.1.2 to define `on' and `off' arm regions,  using the \textit{Spitzer} MIPS 
(Rieke et al. 2004) 24$\mu$m wavelength as a tracer of star formation 
(Calzetti et al. 2007). $24 \mu$m emission is dominated by reprocessed radiation from young
massive stars as demonstrated by its excellent spatial correlation with
other diagnostics associated with such  stars (see
  Relano \& Kennicutt 2009 for such a demonstration in the
case of H$\alpha$ emission in M33).
Helou et al.  (2004) argued that the stellar continuum contributes at most a  few per cent at $24 \mu$m

 In Figure 19 
we plot the ratio of the {\it excess} emission at  24$\mu$m
in the arm compared with the inter-arm region (where the excess is defined
as the mean emission levels in that region minus the minimum emission level
in that radial ring: note that in the case of the $8 \mu$m data this
minimum level is essentially zero in many regions and so in contrast
Figure 18  presents the ratio of unsubtracted values.)

 Figure 19 
is qualitatively similar to Figure 18  and shows a clear dependence of the
star formation response (as measured at $24 \mu$m)  and the amplitude of
variations in the stellar potential.  
  The levels of contrast can be used to interpret 
 Figure 11, noting that the dispersion in specific star formation
rates and in the $24 \mu$m excess ratios are comparable. This suggests 
 that strong spiral structure can contribute
significantly to the dispersion in specific star formation rates in our sample.
Nevertheless, it cannot be the only driver since there are some notable
counter-examples (for example, NGC 3031 with its strong $24 \mu$m excess
ratio and low specific star formation rate. 

\section{Conclusions}\label{discuss}

 By far the strongest correlation that we have identified
(Figure 17) is between
the {\it amplitude} of spiral structure and the strength of tidal
interaction as measured by the $P$ parameter (equation (1)).
We conclude - in line with our conclusions in Paper I - that
a close tidal encounter appears to be a sufficient but not necessary
condition for prominent spiral features. NGC 3198 is a good example
of a galaxy that is unbarred and isolated and yet shows the `grand
design structure' that permits its inclusion in the detailed sample
studied in this paper. Nevertheless, the amplitude of its spiral
features is admittedly lower than the strongly interacting galaxies
in our sample and moreover the $m=2$ mode is less dominant than in these
galaxies.

 We have
indicated the galaxies with the three largest values of $P$ (the
strongly interacting galaxies) by the large crosses   in
all the other correlation plots. These galaxies are  not atypical
for the sample with regard to their pitch angles and the radial
extent of spiral structure (as measured by the azimuthal angle subtended
by the arms in Figures 8 and 9)  and so do not strongly influence these plots.
They are however all rather massive and gas deficient
galaxies of type Sbc or earlier and we must therefore be mindful that
apparent correlations may be driven by this association. This may
be particularly relevant to the interpretation of Figures 4, 5 and 10.

  Our sample covers a rather narrow range of pitch angles and it is thus
unsurprising that we have not identified any correlations between pitch angles
and other galactic parameters. Figure 12 may throw some light on
previous comparisons between the pitch angles of spiral structure in the
optical and NIR: for example in the most recent and sophisticated analysis
of this issue by Davis et al. (2012) there is generally good agreement
between pitch angles in the B and I bands. Discrepant galaxies have a tendency
to have larger pitch angles at longer wavelengths. Figure 12 (which
compares NIR pitch angles with pitch angles measured in H$\alpha$, an
optical  star
formation indicator) suggests that in our sample a discrepancy
of this sign  is evident only for earlier Hubble types.
This would suggest - if borne out in larger samples - that an
association between early Hubble types and tightly wound arms may
only be manifest in the optical (Clarke et al. 2010).   In the most discrepant
galaxies (NGC 3031 and NGC 4579)
 this difference in pitch angles can be explained by our
previous demonstration (Kendall et al. 2008, 2011) that in these galaxies the
spiral pattern traced at $8 \mu$m in shocked gas is significantly
displaced upstream with respect to the NIR (stellar) arms. In
a differentially rotating galaxy it would then be
expected that the spiral traced by
star formation indicators should be more tightly wound than the stellar arms.
The origin of azimuthal offsets is discussed in our previous papers;
it is currently unclear whether there is a more general  association
between such offsets and
galaxies of early Hubble type.

  We also  draw attention to the other strong correlation that we have found, i.e. the demonstration in Figures 18 and 19 of a clear dependence 
 of  the amplitudes of spiral
structure in star formation ($24 \mu$m) and shock ($8 \mu$m) tracers
on the amplitude of potential variations as traced in  the NIR.
 The amplitude of spiral structure at $24 \mu$m demonstrates that
the star formation within the arms cannot on its own account for all  the variation
 in specific star formation rate exhibited by the galaxies in our sample
and there are indeed objects with strong arm contrast yet low specific
star formation rates. However the range of  arm amplitudes 
at $24 \mu$m suggest  that variation of strength of spiral features plays
  a significant contributory 
role 
in setting the specific  star formation rates in disc galaxies.

\section{Acknowledgments.}Many thanks to Bob Carswell for much useful advice regarding the use of IDL, in particular the CURVEFIT program. We are indebted to Hans-Walter Rix for the idea of using colour-correction to make mass maps from optical data. Many thanks also go to Jim Pringle, Clare Dobbs,  Stephanie Bush,  
Giuseppe Bertin and Jerry Sellwood for useful discussions over the course of this work and to Phil James and Marc Seigar for valuable observational input. 
We also thank the referee for constructive comments. 
This work makes use of IRAF. IRAF is distributed by the National Optical Astronomy Observatories, which are operated by the Association of Universities for Research in Astronomy, Inc., under cooperative agreement with the National Science Foundation.


\label{lastpage}

\end{document}